\documentclass[twocolumn,conference]{IEEEtran}
\usepackage[T1]{fontenc}
\usepackage[latin9]{inputenc}
\usepackage{color}
\usepackage{verbatim}
\usepackage{amsmath}
\usepackage{graphicx}
\usepackage[unicode=true,
 bookmarks=true,bookmarksnumbered=true,bookmarksopen=true,bookmarksopenlevel=1,
 breaklinks=false,pdfborder={0 0 0},pdfborderstyle={},backref=false,colorlinks=false]
 {hyperref}
\hypersetup{pdftitle={Your Title},
 pdfauthor={Your Name},
 pdfpagelayout=OneColumn, pdfnewwindow=true, pdfstartview=XYZ, plainpages=false}
\usepackage{breakurl}

\makeatletter
\usepackage[caption=false,font=footnotesize]{subfig}
\usepackage{amsmath}

\makeatother

\begin{document}

\title{On the Interplay Between Edge Caching and HARQ in Fog-RAN}

\author{\IEEEauthorblockN{Igor Stanojev\IEEEauthorrefmark{1} and Osvaldo Simeone\IEEEauthorrefmark{2}}\IEEEauthorblockA{\IEEEauthorrefmark{1}University of Wisconsin-Platteville, Platteville,
USA}\IEEEauthorblockA{\IEEEauthorrefmark{2}King's College London, London, UK}}
\maketitle
\begin{abstract}
In a Fog Radio Access Network (Fog-RAN), edge caching is combined
with cloud-aided transmission in order to compensate for the limited
hit probability of the caches at the base stations (BSs). Unlike the
typical wired scenarios studied in the networking literature in which
entire files are typically cached, recent research has suggested that
fractional caching at the BSs of a wireless system can be beneficial.
This paper investigates the benefits of fractional caching in a scenario
with a cloud processor connected via a wireless fronthaul link to
a BS, which serves a number of mobile users on a wireless downlink
channel using orthogonal spectral resources. The fronthaul and downlink
channels occupy orthogonal frequency bands. The end-to-end delivery
latency for given requests of the users depends on the HARQ processes
run on the two links to counteract fading-induced outages. An analytical
framework based on theory of Markov chains with rewards is provided
that enables the optimization of fractional edge caching at the BSs.
Numerical results demonstrate meaningful advantages for fractional
caching due to the interplay between caching and HARQ transmission.
The gains are observed in the typical case in which the performance
is limited by the wireless downlink channel and the file popularity
distribution is not too skewed.

\end{abstract}

\section{Introduction}

In recent years, the placement of caches in communication networks
has progressively moved from the the Internet-located data centers
of Content Delivery Networks to the core or access network of Internet
Service Providers (as in Netflix Open Connect). The logical end point
of this trend is edge caching, or femto-caching, that is, the storage
of popular content directly at the Base Stations (BSs) \cite{Spectr_effic}.
While initial work in the networking literature on the subject provided
discouraging results due to low hit rates at the BSs, more recent
research has argued that the rapid decrease of the cost of storage
makes edge caching a potentially desirable technology \cite{Spectr_effic}.%
\begin{comment}
In order to accommodate a broad range of communication services that
are envisioned to be within the scope of 5G systems and beyond, a
novel architecture Cloud Radio Access Network (C-RAN) recently emerged,
prescribing the virtualization of baseband functionalities from the
base stations (BS) to the cloud processor \cite{Cloud-RAN}-\cite{C-RAN Osv}.
While this approach can increase spectral efficiency \cite{Spectr_effic},
it comes on the account of delay, due to the increased amount of data
communicated on the fronthaul link between the cloud and BS \cite{Fog-RAN}.
To address this drawback, a hybrid architecture Fog-RAN (F-RAN) has
been recently advocated, wherein the BS may be endowed with caching
capabilities, in order to respond with low latency to local data requests
and to alleviate the fronthaul traffic \cite{Fog-RAN}\cite{key-2 Fog}.
\end{comment}

The vast literature on the topic of cache management in wired content
delivery networks by and large assumes the indivisibility of each
content in the library and focuses on the design of online content
replacement strategies under dynamic models for the content requests,
see, e.g., \cite{key-1 ICN}. Furthermore, initial works on femto-caching
such as \cite{Spectr_effic} are also based on the assumption of indivisible
contents, as well as on a simplified modeling of the wireless channels
in terms of coverage areas.%
\begin{comment}
The capacity of data repositories at BS, i.e., the caches, will be
able to accommodate only a small fraction of the plethora of data
that can be requested by a mobile user (MU) \cite{key-2 Fog}. Thus,
only a fraction of the most popular data will be cached at BS, while
the remainder must still be fetched from the cloud via fronthaul \cite{key-2 Fog}\cite{key-3 Big Data}.
In this paper, we consider caching policies that allow not only for
caching entire files, but also for caching of fractions of files,
similarly as in \cite{Fog-RAN}. With such a policy, a larger set
of files (their fractions) can be cached, and the network can respond
faster to a larger number of MU's requests. While the policy implies
that the remaining fraction of the requested file has to be transmitted
on the fronthaul link from the cloud to BS, and then in the downlink
from BS to a MU, the consequent increase of the delay can be minimized
by leveraging a pipelined transmission of the cached and cloud data
on the two links.
\end{comment}

\begin{figure}[tbh]
\begin{centering}
\includegraphics[scale=0.53]{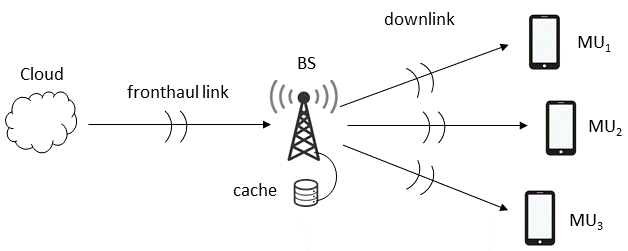}
\par\end{centering}
\centering{}\caption{System model with three MUs. \label{fig:System-Model}}
\end{figure}
In more recent research, starting with \cite{key-8 Maddah}, the interplay
of interference management and edge caching was studied by accounting
for the superposition and broadcast properties of wireless transmission.
This line of work, including \cite{key-10 Sengupta}-\cite{key-9 Cache aided interference mngmnt}
among others, concentrates on the high-signal to noise ratio (SNR)
regime. A main conclusion from these papers is that fractional caching
at the BSs of a wireless system can be beneficial in terms of number
of achievable degrees of freedom. The key reason for the potential
gain of fractional caching is the enhanced flexibility afforded by
fractional caching in enabling coordinated transmission at distributed
BSs.

Despite the improvements in storage technologies, it is still expected
that the capacity of caches at the BSs will be able to accommodate
only a small, though possibly not negligible, fraction of the contents
that may be requested by mobile users (MUs) (see, e.g., \cite{key-2 Fog}
\cite{key-3 Big Data}). Thus, the uncached requested contents will
have to be fetched from a content provider via fronthaul or backhaul
links so as to be available at the BSs for delivery. Based on this
observation, references \cite{key-10 Sengupta}, \cite{Tandon}-\cite{Azimi ISIT},
studied a more general set-up, which includes not only edge caching
but also a cloud processor with access to the content provider. In
this class of systems referred to as Fog Radio Access Networks (Fog-RANs),
the cloud processor is connected to the BSs via fronthaul links that
can be used to deliver uncached information. References \cite{key-10 Sengupta}
\cite{Tandon} \cite{Azimi ISIT} demonstrated the advantages of fractional
caching in this scenario and the dependence of the optimal caching
strategy on the fronthaul capacity. %
\begin{comment}
The paper investigates the optimal caching strategies that would yield
a minimum delay in the network\textquoteright s response to MU\textquoteright s
data request. In particular, for a set of popular files, where each
file is associated with some probability of request, we search for
the optimum cached fractions for each of these files that would minimize
the data delivery delay, given the limited cache size as a constraint.
In the considered scenario, both fronthaul and downlink are modeled
as wireless links, and the delay is caused by fading-induced link
outages, requiring data retransmissions. We also provide performance
comparisons between the caching of a fraction of whole popular files,
and caching of fractions of each of the popular files (notice that
the former is a special case of the latter). In particular, we find
that the proposed fractional file caching is invaluable when the file
request probability distribution tends to be uniform. ... One of the
advantages of this model is that cloud can multicast messages to multiple
BSs \cite{key-5 Limits}.
\end{comment}

This paper investigates the benefits of fractional caching in a simple
scenario with a cloud processor connected via a wireless fronthaul
link to a BS, which serves a number of mobile users on a wireless
downlink channel using orthogonal spectral resources, as seen in Figure
\ref{fig:System-Model}. Unlike the prior works described above, here
we model the impact of Hybrid Automatic Repeat reQuest (HARQ) processes
run on the two links to counteract fading-induced outages. The driving
question of this work is: \emph{Can fractional caching be advantageous
even in the absence of BS coordination?} We answer this question in
the positive, demonstrating the interplay between edge caching and
HARQ. This is done by deriving an analytical framework based on theory
of Markov chains with rewards that enables the minimization of the
end-to-end latency over fractional edge caching. It is noted that
the effect of retransmissions on caching design was also considered
in \cite{Retransmissions in caching}. Therein, an MU is served by
different BSs, each caching entire files, as it roams the cells. If
the requested file is not cached at the BSs currently serving MU,
or if it is transmitted with an error, the MU will require a retransmission
at a later time when it will possibly be served by a different set
of BSs. We emphasize that \cite{Retransmissions in caching} considers
neither fractional file caching nor fronthaul transmission.%
\begin{comment}
It is noted that the effect of retransmissions on caching design was
also considered in \cite{Retransmissions in caching}. Therein, the
MU is served by a different set of BSs at a time as it roams around
the cells. The BSs avail a different cache of whole files. If the
requested file is not cached at the BSs currently serving MU, or if
it is transmitted with an error, the MU will require a retransmission
in the next moment, when it will possibly be served by a different
set of BSs. Notice that \cite{Retransmissions in caching} does not
consider fractional file caching, neither it utilizes the fronthaul,
and is thus of fundamentally different nature than the work at hand.
\end{comment}

The paper is organized as follows. In Section \ref{sec:System-Model}
we present the system model. Section \ref{sec:Analysis-of-Optimal}
provides details of the analysis and optimization. Numerical insights
can be found in Section \ref{sec:Numerical-Results}, and the paper
is concluded in Section \ref{sec:Conclusions}.

\section{System Model\label{sec:System-Model}}

We consider a downlink transmission model consisting of a cloud processor,
a BS, and a number of MUs, as illustrated in Figure \ref{fig:System-Model}.
The cloud is connected to the BS via a wireless fronthaul link, while
the BS communicates with the MUs over a wireless downlink channel
using orthogonal spectral resources. The operation of a system is
divided into a placement, or caching, phase and a delivery phase,
as in most related papers, e.g., \cite{key-8 Maddah} \cite{key-10 Sengupta}
\cite{Tandon} \cite{key-5 Limits}.

We assume that there are $F$ popular files and that each file can
be split into $N$ packets, each to be transmitted in a separate physical
layer frame. All files are available in the cloud. During the placement,
or caching, phase, $N_{f}$ packets of any popular file $f$ are stored
in the BS's cache, where $N_{f}\in\left\{ 0,1,..,N\right\} $ and
$f\in\left\{ 1,2,..,F\right\} $. The $F$ parameters $N_{f}$ will
be the subject to optimization in Section \ref{sec:Analysis-of-Optimal}
under a cache capacity constraint $\sum_{f=1}^{F}N_{f}\leq C,$ where
$C$ is the BS's cache capacity in numbers of packets.

In the delivery phase, the BS serves one MU at a time. Each MU requests
a file $f$ with probability $u_{f}$, with $\sum_{f=1}^{F}u_{f}=1$.
Requests are independent and the probability $u_{f}$ follows the
Zipf distribution (e.g., \cite{Content Centric (zipf)}): 
\begin{equation}
u_{f}=cf^{-\gamma},\text{ \ensuremath{f\in\left\{ 1,2,..,F\right\} }},
\end{equation}
where $\gamma\ge0$ is a given popularity exponent and $c>0$ is the
normalizing constant. Notice that $\gamma=0$ yields a uniform popularity
distribution, while with larger $\gamma$ the distribution becomes
more skewed, with files $f=1,..,F,$ sorted by descending popularity. 

The fronthaul and downlink wireless links use separate frequency bands
of the same size and are frame synchronous, so that each packet transmission
slot, or frame, can accommodate two simultaneous transmissions, one
on each link. The two links are modeled as block-fading Rayleigh channel
gains, with independent zero-mean unit-power complex Gaussian channel
gains, which are constant during each transmission slot and change
independently with each (re)transmission. %
\begin{comment}
The independent zero-mean unit-power complex Gaussian channels are
denoted as $h_{1}^{t}$ and $h_{2}^{t}$ for fronthaul and downlink,
respectively, where $t$ denotes the transmission slot. 
\end{comment}
The average signal-to-noise ratios (SNR) on the two links are denoted
as $\mathrm{SNR}_{1}$ and $\mathrm{SNR_{2}}$. The packet transmission
rate in bits per second per hertz (bit/s/Hz) is denoted as $r$. We
consider HARQ Type I protocol, whereby erroneously received packets
are discarded at the destination. All signaling messages, such as
ACK and NACK messages, are assumed to be significantly shorter than
the user data packets and to be transmitted with perfect reliability.

The fraction of users at a given distance $d$ from the BS is evaluated
by assuming a uniform MU's placement distribution within a circular
cell of radius $R$. This fraction is proportional to distance $d$
and reads 
\begin{equation}
v(d)=\frac{2d}{R^{2}},\:0\le d\le R.
\end{equation}
We note that any other distribution could be accommodated in the analysis
and that further details will be provided in Section \ref{subsec:Caching-Optimization}.
The average downlink signal to noise ratio $\mathrm{SNR_{2}}$ follows
the path-loss model
\begin{equation}
\mathrm{SNR_{2}}(d)=\frac{K}{d^{\mu}},
\end{equation}
where $\mu$ is the propagation-loss exponent, and $K$ is a constant
that depends on the transmission power of the BS and that sets the
signal-to-noise ratio $\mathrm{SNR_{2}}$ at $d=1$ m.

As the performance metric, we use end-to-end average delay, i.e.,
the average number of transmission slots required to deliver all $N$
packets of a requested file to all the MUs.

\section{Optimal Caching Policy\label{sec:Analysis-of-Optimal}}

In this section, we first analyze the impact of the number $N_{f}$
of cached packets on the delay for a single MU requesting file $f$
in Section \ref{subsec:Delay-Analysis-For}. Then, in Section \ref{subsec:Caching-Optimization},
we incorporate multiple MUs and tackle the problem of optimizing the
cache allocation to minimize the average delivery latency.

\subsection{Delay Analysis For a Given File\label{subsec:Delay-Analysis-For}}

Here, we evaluate the transmission delay as a function of the number
$N_{f}$ of cached packets for a given requested file $f$. The probability
of successful transmission, i.e., the probability that a retransmission
is not required, is the probability that the channel capacity can
accommodate a transmission of rate $r$. This can be found to be (e.g.,
\cite{key-2 HARQ}):
\begin{align}
p_{l} & =e^{-\frac{2^{r}-1}{2\mathrm{SNR}_{l}}},\text{ \ensuremath{l\in\left\{ 1,2\right\} }}
\end{align}
where indices $l=1$ and $l=2$ identify the probability of successful
transmission on fronthaul and downlink, respectively.

\begin{figure}[tbh]
\begin{centering}
\includegraphics[scale=0.24]{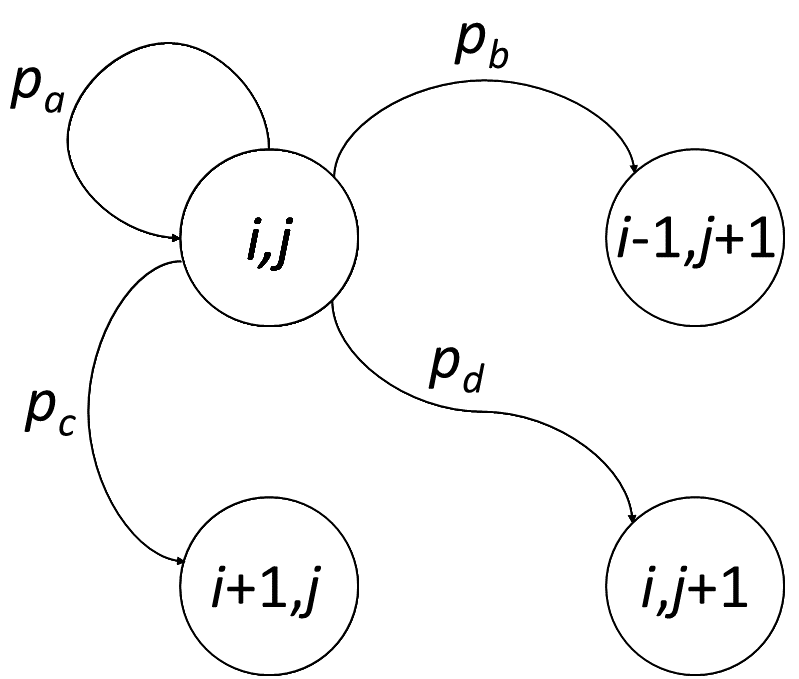}
\par\end{centering}
\caption{Illustration of the outgoing state transitions for a non-sink state
$(i,j)$ for the Markov chain analyzed in Section \ref{subsec:Delay-Analysis-For}.\label{fig:Detail-of-the}}
\end{figure}
In order to evaluate the end-to-end average delay, we use a Markov
chain analysis. Towards this goal, we define the state of the Markov
chain as the pair $(i,j)$, where $i=0,1,..,N,$ is the number of
packets at the BS that are yet to be delivered to the MU, and $j=0,1,..,N$
is the number of packets already delivered to the MU. Note that a
state $(i,j)$ is admissible only if the inequalities $N_{f}\leq i+j\leq N$
are satisfied. The initial state is $(N_{f},0)$, while the absorbing,
or sink, state is $(0,N)$. Transitions from a state $(i,j)\neq(0,N)$
are shown in Figure \ref{fig:Detail-of-the}. The transition probabilities
follow from the description of the system model and can be derived
as\begin{subequations}
\begin{align}
p_{a} & =\begin{cases}
1-p_{2}, & \text{if }i+j=N\\
1-p_{1}, & \text{if }i=0\\
\left(1-p_{1}\right)\left(1-p_{2}\right), & \text{otherwise}
\end{cases}\\
p_{b} & =\begin{cases}
p_{2}, & \text{if }i+j=N\\
0, & \text{if }i=0\\
\left(1-p_{1}\right)p_{2}, & \text{otherwise}
\end{cases}\\
p_{c} & =\begin{cases}
0, & \text{if }i+j=N\\
p_{1}, & \text{if }i=0\\
p_{1}\left(1-p_{2}\right), & \text{otherwise}
\end{cases}\\
\text{and\thinspace\ }p_{d} & =\begin{cases}
0, & \text{if }i+j=N\text{ or \ensuremath{i=0}}\\
p_{1}p_{2}, & \text{otherwise.}
\end{cases}\label{eq:p_wrong}
\end{align}
\label{eq:p_abcd}\end{subequations}Notice that the conditions $i+j=N$
denotes the event that there is no transmission on the fronthaul since
the BS has received all $N-N_{f}$ packets from the cloud, while the
event $i=0$ indicates that there is no transmission on the downlink
given that the MU has received all packets currently available at
BS. 

To compute the average end-to-end delay, i.e., the average number
of transmission slots needed for the complete delivery of $N$ packets
to the MU, we apply the theory of Markov chains with rewards {[}16,
Chapter 4{]}. Denote the average number of transmission slots, or
steps of the Markov chain, required to reach the sink state from a
state $(i,j)$ as $\nu_{i,j}$. These can be obtained from Figure
\ref{fig:Detail-of-the} and (\ref{eq:p_abcd}) as:
\begin{equation}
\nu_{i,j}=\begin{cases}
0, & \!\!\!\!\!\!\!\!\!\!\!\!\!\!\!\!\!\!\!\!\!\!\!\!\!\!\!\!\!\!\!\!\!\!\!\!\!\!\!\!\!\!\!\!\!\!\!\!\!\!\!\!\!\!\!\!\!\!\!\!\!\!\!\!\!\!\!\!\text{for }\left(i=0,j=N\right)\text{ (sink)}\\
1+(1-p_{2})\nu_{i,j}+p_{2}\nu_{i-1,j+1}, & \!\!\!\!\!\!\!\!\!\!\!\!\!\!\!\!\!\!\!\!\!\!\!\!\!\!\!\!\!\!\!\!\!\!\!\!\!\!\!\text{for }i+j=N\\
1+(1-p_{1})\nu_{i,j}+p_{1}\nu_{i,j+1}, & \!\!\!\!\!\!\!\!\!\!\!\!\!\!\!\!\!\!\!\!\!\!\!\!\!\!\!\text{for }i=0\\
1+(1-p_{1})(1-p_{2})\nu_{i,j}+(1-p_{1})p_{2}\nu_{i-1,j+1}\\
+p_{1}\left(1-p_{2}\right)\nu_{i+1,j}+p_{1}p_{2}\nu_{i,j+1}, & \!\!\!\!\!\!\!\!\!\!\!\!\!\!\!\!\!\!\!\!\!\!\!\!\!\!\!\!\!\text{otherwise.}
\end{cases}\label{eq: nu_ij}
\end{equation}
For each of the four cases in (\ref{eq: nu_ij}), the parameter $\nu_{i,j}$
can be expressed explicitly, yielding:
\begin{equation}
\nu_{i,j}=\begin{cases}
0, & \!\!\!\!\!\!\!\!\!\!\!\!\!\!\!\!\!\!\!\!\!\!\!\!\!\!\!\!\!\!\!\!\!\!\!\!\!\!\!\!\!\!\!\!\!\!\!\!\!\text{for }\left(i=0,j=N\right)\text{ (sink)}\\
\frac{1+p_{2}\nu_{i-1,j+1}}{p_{2}}, & \!\!\!\!\!\!\!\!\!\!\!\!\!\!\!\!\!\!\!\!\text{for }i+j=N\\
\frac{1+p_{1}\nu_{i+1,j}}{p_{1}}, & \!\!\!\!\!\!\!\!\text{for }i=0\\
\frac{1+(1-p_{1})p_{2}\nu_{i-1,j+1}+p_{1}(1-p_{2})\nu_{i+1,j}+p_{1}p_{2}\nu_{i,j+1}}{1-\left(1-p_{1}\right)\left(1-p_{2}\right)}, & \!\!\!\text{\!\!\!otherwise.}
\end{cases}\label{eq:nu_i,j set}
\end{equation}
This set can be easily solved recursively, starting from the sink
state $(0,N)$ and moving backwards towards the initial state $(N_{f},0)$. 

The average end-to-end delay is then equal to the average number of
steps required to reach the sink from the initial state $\left(N_{f},0\right)$,
i.e.,
\begin{equation}
T_{N_{f}}(\mathrm{SNR_{2}})=\nu_{N_{f},0}.\label{eq:T_dep}
\end{equation}
In the notation adopted in (\ref{eq:T_dep}), we emphasized that the
derived delay depends on the number $N_{f}$ of cached packets for
the requested file $f$ and on the average signal-to-noise ratio $\mathrm{SNR_{2}}$
of the MU.

\subsection{Caching Optimization\label{subsec:Caching-Optimization}}

In this section, we use the result (\ref{eq:T_dep}) to tackle the
optimization of the cache allocation variables $N_{f}$, $f\in\left\{ 1,2,..,F\right\} $.
We discretize the BS-MU distances to a set of $k$ distances $d_{i}=i/k\cdot R$,
$i=1,..,k$. Assuming a random and uniform placement of MUs in a circular
cell, the fraction of MUs at distance $d_{i}$ is given by
\begin{equation}
v_{i}=\frac{2i}{k(1+k)},\text{ \ensuremath{i=1,..,k}}.
\end{equation}

The average end-to-end delay is obtained by averaging (\ref{eq:T_dep})
with respect to the MU distances from the BS, and over the files popularity
distribution, yielding:
\begin{equation}
T=\sum_{i=1}^{k}v_{i}\sum_{f=1}^{F}u_{f}T_{N_{f}}\left(\mathrm{SNR_{2}}(d_{i})\right).\label{eq:T_final}
\end{equation}
We are interested in minimizing the average delay $T$ in (\ref{eq:T_final})
over the caching policy, i.e., over the variables $N_{f}$, $f\in\left\{ 1,2,..,F\right\} $.
For this purpose, let us introduce the binary indicator optimization
variables $q_{fn}$ for each file $f$, where $n\in\left\{ 0,1,...,N\right\} $:
\begin{equation}
q_{fn}=\begin{cases}
1, & \text{if }n=N_{f}\\
0, & n\in\left\{ 0,1,..,N\right\} \setminus\left\{ N_{f}\right\} 
\end{cases}\label{eq:indicator}
\end{equation}
With this definition, the number $N_{f}$ of cached packets and the
delay $T_{N_{f}}$ can be expressed as $N_{f}=\sum$$_{n=0}^{N}nq_{fn}$
and $T_{N_{f}}=\sum$$_{n=0}^{N}q_{fn}T_{n}$, respectively, for $f\in\left\{ 1,2,..,F\right\} $.
Furthermore, the optimization can be formulated as:\begin{subequations}
\begin{gather}
\text{min \ensuremath{\sum_{i=1}^{k}\sum_{f=1}^{F}\sum_{n=0}^{N}v_{i}u_{f}q_{fn}T_{n}\left(\mathrm{SNR_{2}}(d_{i})\right)}}\text{}\label{eq: wrong}\\
\!\!\!\!\!\!\!\!\!\!\!\!\!\!\!\!\!\!\!\!\!\!\!\!\!\!\!\!\!\!\!\!\!\!\!\!\!\!\!\!\!\!\!\!\!\!\!\!\!\!\!\!\!\!\!\!\!\text{s.t. }q_{fn}\in\left\{ 0,1\right\} \label{eq:ineq1}\\
\!\!\!\!\!\!\!\!\!\sum_{n=0}^{N}q_{fn}\leq1,\text{ \ensuremath{f\in\left\{ 1,..,F\right\} }}\label{eq:ineq2}\\
\!\!\!\!\!\!\!\!\!\!\!\!\!\!\!\!\!\!\!\!\!\!\!\!\!\!\!\!\!\!\sum_{f=1}^{F}\sum_{n=0}^{N}nq_{fn}\leq C.\label{eq:ineq3}
\end{gather}
\label{eq: optimization}\end{subequations}The inequalities (\ref{eq:ineq1})-(\ref{eq:ineq2})
impose that, for a particular file $f,$ exactly one of the indicators
$q_{fn}$, $n\in\left\{ 0,1,..,N\right\} $, equals to one, while
the others must be zero (recall (\ref{eq:indicator})). The inequality
(\ref{eq:ineq3}) enforces the cache capacity constraint. The problem
(\ref{eq: optimization}) is a linear integer (binary) optimization
problem, a class of optimization problems which can be solved using
readily available fast algorithms \cite{Integer}.

\section{Numerical Results\label{sec:Numerical-Results}}

In this section, we provide insights into the interplay between edge
caching and HARQ retransmissions via numerical results. Throughout,
we set the file size to $N=20$ packets, and packet transmission rate
to $r=2$ bit/s/Hz.

\begin{figure}[tbh]
\centering{}\includegraphics[scale=0.38]{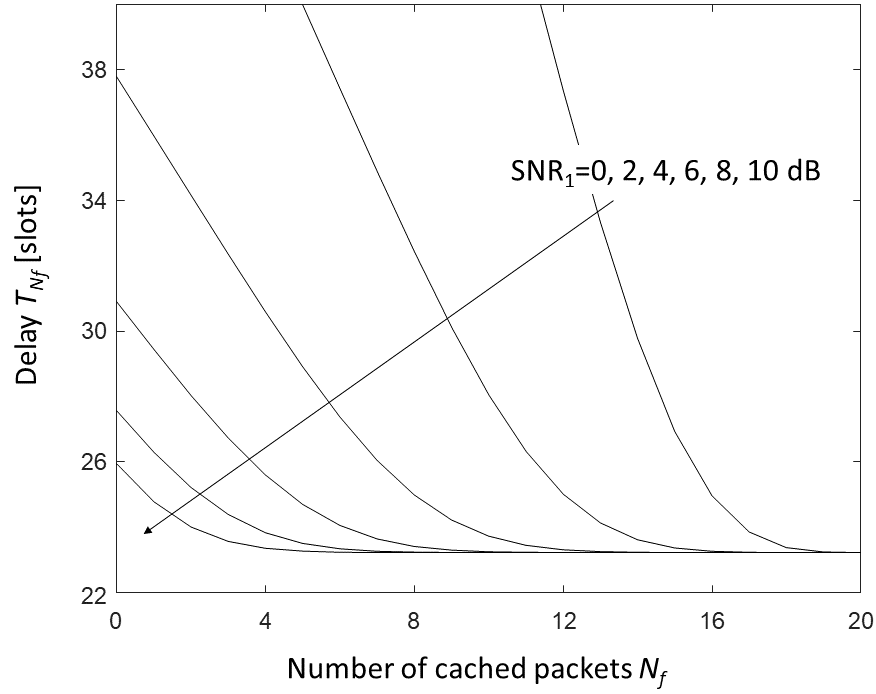}\caption{End-to-end average delay for a single user as a function of the number
$N_{f}$ of cached packets for the requested file $f$, versus the
fronthaul SNR $\mathrm{SNR_{1}}$ ($\mathrm{SNR_{2}}=10$ dB, $r=2$
bit/s/Hz, $N=20$ packets (file size)).\label{fig:Delay1}}
\end{figure}
We start by investigating the impact of the number of cached packets
on the average delay for a given file request, as detailed in Section
\ref{subsec:Delay-Analysis-For}. Figure \ref{fig:Delay1} shows the
delay $T_{N_{f}}$ in slots (packet transmissions) as a function of
the number of cached packets $N_{f}$ for different values of the
fronthaul SNR $\mathrm{SNR_{1}}$ with $\mathrm{SNR_{2}}=10$ dB.
As a first remark, the average delay cannot drop below approximately
23 slots, which is the minimum dictated by the downlink retransmissions.
It can also be observed that the larger the fronthaul SNR $\mathrm{SNR_{1}}$
is, the smaller is the cache capacity, measured here by $N_{f}$,
necessary to reach the minimum average delay. Namely, when $\mathrm{SNR_{1}}$
is large, the uncached packets can be fetched from the cloud on the
fronthaul during downlink transmission. This shows that fractional
caching, as opposed to the full caching of files, typically assumed
in the networking literature (see, e.g., \cite{key-9 Cache aided interference mngmnt}),
can be implemented without loss of optimality in the presence of HARQ.
A similar observation was made in \cite{key-7 Leconte}. 

\begin{figure}[tbh]
\begin{centering}
\includegraphics[scale=0.38]{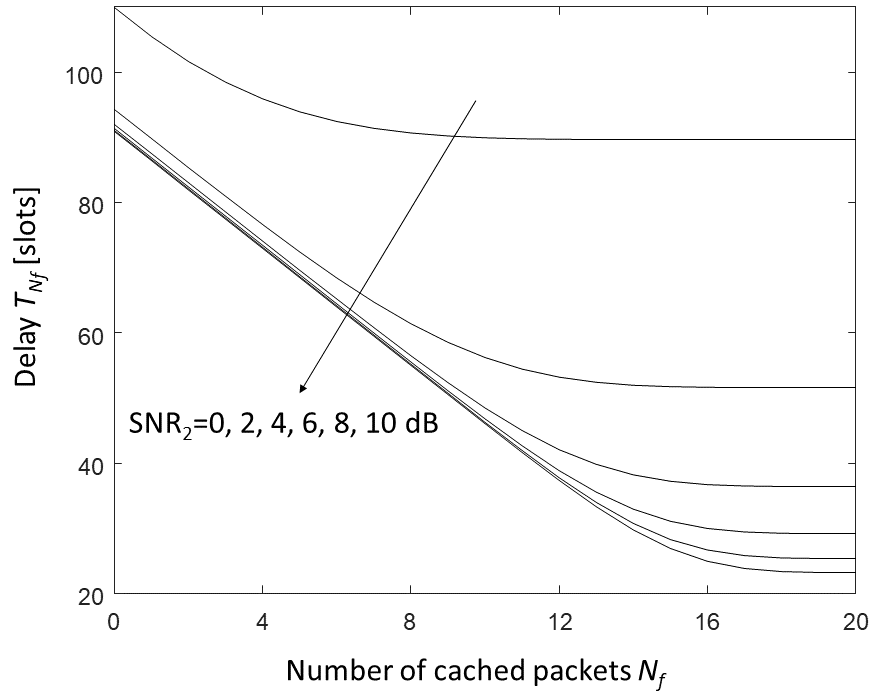}
\par\end{centering}
\caption{End-to-end average delay for a single user as a function of the number
$N_{f}$ of cached packets for the requested file $f$, versus the
downlink SNR $\mathrm{SNR_{2}}$ ($\mathrm{SNR_{1}}=0$ dB, $r=2$
bit/s/Hz, $N=20$ packets (file size)).\label{fig:Delay2}}
\end{figure}
In Figure \ref{fig:Delay2}, we study the effect of the downlink SNR
$\mathrm{SNR_{2}}$ on the end-to-end average delay by showing the
delay $T_{N_{f}}$ as a function of the number of cached packets $N_{f}$,
with $\mathrm{SNR_{1}}=0$ dB. In order to obtain the minimum delay
for a given value of $\mathrm{SNR_{2}}$, a larger number of cached
packets is required for larger values of $\mathrm{SNR_{2}}$ so as
to compensate for the lower fronthaul SNR. Another observation is
that, for a small number of cached packets $N_{f}$, the delay decreases
linearly with $N_{f}$ and does not decrease significantly with the
increase of $SNR_{2}$, as it is dominated by the quality of the fronthaul
link. 

\begin{figure}[tbh]
\begin{centering}
\includegraphics[scale=0.38]{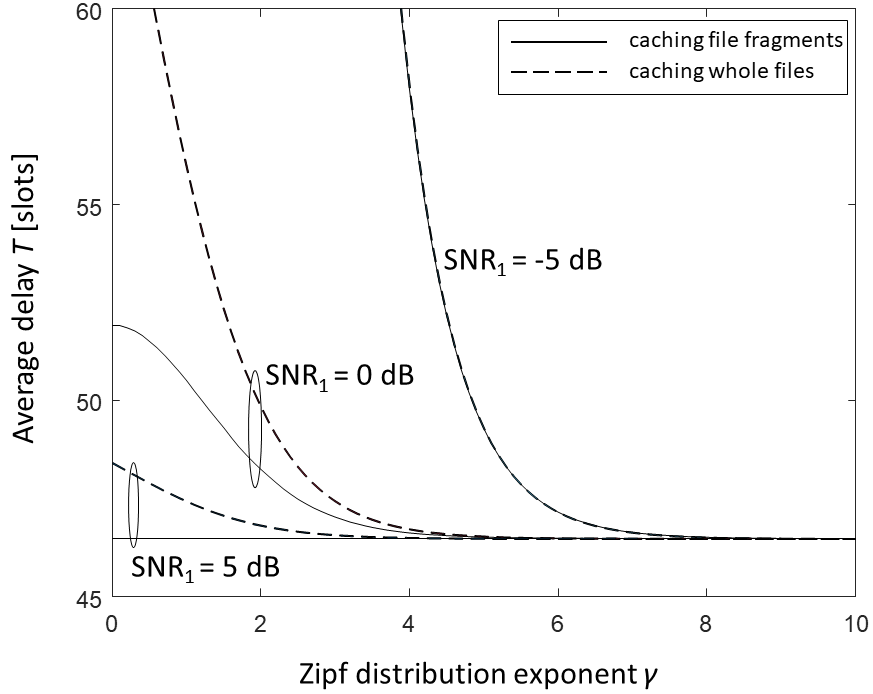}
\par\end{centering}
\caption{End-to-end average delay under optimum caching policy, as a function
of the Zipf exponent $\gamma$, versus the fronthaul SNR $\mathrm{SNR_{1}}$
($r=2$ bit/s/Hz, $N=20$ packets (file size), $C=60$ packets (cache
capacity), $F=5$ files, $R=100$ m, $K=40$ dB, $\mu=2$).\label{fig:Delay3g}}
\end{figure}
We now turn to the end-to-end average delay under the optimum caching
policy discussed in Section \ref{subsec:Caching-Optimization}, while
accounting for the random placement of users in the cell. In Figure
\ref{fig:Delay3g}, we show the minimum average delay as a function
of the Zipf exponent $\gamma$ for different values of the fronthaul
SNR $\mathrm{SNR_{1}}$, with $F=5$ files and a cache capacity equal
to $C=60$ packets (i.e., three files). Additionally, the cell range
is taken as $R=100$ m, with the user distance discretized to $k=1000$
values, and we set $K=40$ dB (recall that $K$ is the value of $\mathrm{SNR_{2}}$
at distance $d=1$ m), and propagation factor $\mu=2$ (this yields
11.7 dB for the average $\mathrm{SNR_{2}}$). Figure \ref{fig:Delay3g}
also compares the delay under the optimum caching policy derived in
Section \ref{subsec:Caching-Optimization} whereby file fractions
can be cached, with the more conventional set-up where only whole
files can be cached. As expected, the former performs at least as
well as the latter. More interestingly, the performance of the two
policies is identical for larger values of $\gamma$, i.e., for a
skewed Zipf distribution. In this regime, the popularity of files
is concentrated on the more popular files and it is optimal to cache
the complete most popular files. A general conclusion here is that
the design degree of freedom to cache fractions of files is beneficial
in presence of a more uniform popularity distribution (i.e., a small
$\gamma$).

\begin{figure}[tbh]
\begin{centering}
\includegraphics[scale=0.38]{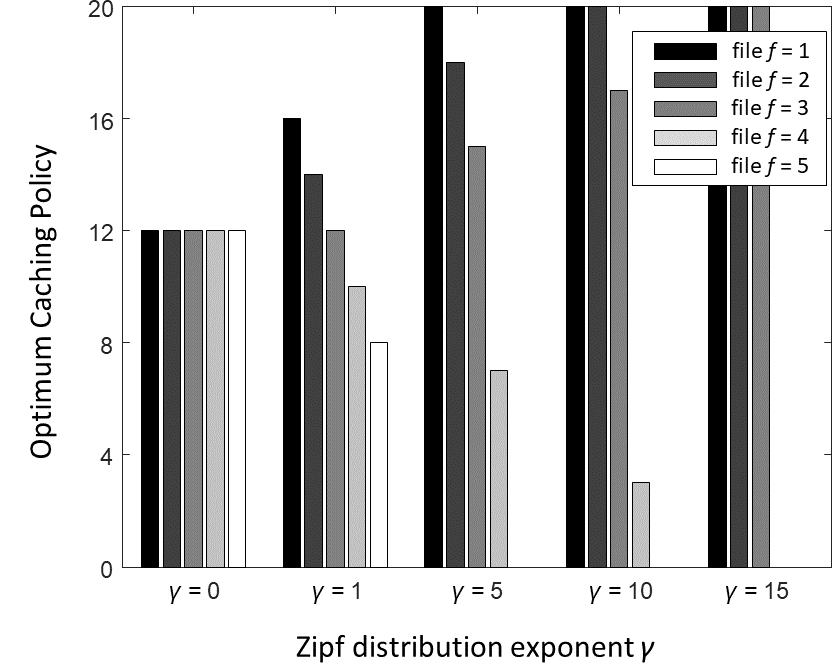}
\par\end{centering}
\caption{Optimum caching policy as a function of the Zipf exponent $\gamma$
($\mathrm{SNR_{1}}=0$ dB, $r=2$ bit/s/Hz, $N=20$ packets (file
size), $C=60$ packets (cache capacity), $F=5$ files, $R=100$ m,
$K=40$ dB, $\mu=2$).\label{fig:Optimum-Caching-Policy}}
\end{figure}
This point is corroborated in Figure \ref{fig:Optimum-Caching-Policy},
which presents the optimum caching policies for different values of
the Zipf exponent $\gamma$, when $\mathrm{SNR_{1}}=0$ dB, and for
the remaining parameters as in Figure \ref{fig:Delay3g}. The optimum
caching policy is to cache equal fractions of each file when the file
popularity distribution is uniform ($\gamma=0$), while with the increase
of $\gamma$, the caching policy starts to resemble the conventional
one of caching the whole files.

\section{Conclusions\label{sec:Conclusions}}

The main conclusion of this work is that caching fractions of files
at a BS can significantly improve over the standard approach of caching
entire files when the performance is limited by the wireless downlink
channel and the file popularity distribution is not too skewed. This
is due to the interplay between fractional caching and HARQ: as the
BS performs retransmissions on the downlink channel to ensure reliable
communication, the fronthaul link can deliver uncached portions of
a file. Interesting open aspects include the investigation of the
interaction between the gains identified here and the benefits due
to cooperation in the presence of multiple BSs studied in \cite{key-10 Sengupta}.

\section{Acknowledgment}

The work of O. Simeone was partially supported by the U.S. NSF through
grant CCF-1525629. O. Simeone has also received funding from the European
Research Council (ERC) under the European Union\textquoteright s Horizon
2020 research and innovation programme (grant agreement No 725731).

\end{document}